# Optimizing Thermal Transport in Graphene Nanoribbon Based on Phonon Resonance Hybridization

Xiao Wan[1,2#], Dengke Ma[3#], Dongkai Pan[1,2], Lina Yang[4], Nuo Yang[1,2]*

1. State Key Laboratory of Coal Combustion, Huazhong University of Science and Technology, Wuhan 430074, China.
2. School of Energy and Power Engineering, Huazhong University of Science and Technology, Wuhan 430074, China.
3. NNU-SULI Thermal Energy Research Center (NSTER) & Center for Quantum Transport and Thermal Energy Science (CQTES), School of Physics and Technology, Nanjing Normal University, Nanjing 210023, China.
4. School of Aerospace Engineering, Beijing Institute of Technology, Beijing 100081, China.

# X.W. and D.M. contributed equally to this work.
*Corresponding email: nuo@hust.edu.cn (N.Y)

# Abstract

As a critical way to modulate thermal transport in nanostructures, phonon resonance hybridization has become an issue of great concern in the field of phonon engineering. In this work, we optimized phonon transport across graphene nanoribbon and obtained minimized thermal conductance by means of designing pillared nanostructures based on resonance hybridization. Specifically, the optimization of thermal conductance was performed by the combination of atomic Green's function and Bayesian optimization. Interestingly, it is found that thermal conductance decreases non-monotonically with the increasing of number for pillared structure, which is severed as resonator and blocks phonon transport. Further mode-analysis and atomic Green's function calculations revealed that the anomalous tendency originates from decreased phonon transmission in a wide frequency range. Additionally, nonequilibrium molecular dynamics simulations are performed to verify the results with the consideration of high-order phonon scattering. This finding provides novel insights into the control of phonon transport in nanostructures.

# 1. Introduction

Graphene nanoribbons (GNRs) have attracted great attention recently due to its extraordinary properties and potential applications in various fields [1,2]. One promising application relates to thermolelectrics [3,4]. Although the thermal conductivity of GNR is high, its mechanical flexibility and outstanding electric properties, give it potential to be applied in thermoelectrics [4]. The ability to reduce the thermal conductivity of GNR without degrading the electric properties provides a practicable avenue for improving thermoelectric performance. Besides, it's essential for understanding heat transfer physics in low-dimensional system as well.

Tailoring thermal conductivity of materials in atomic scale through phonon engineering is a rapidly growing area of research [5-12]. Different types of nanostructures have been proposed to reduce thermal conductivity of GNRs based on the particle nature of phonon, such as vacancies [13], edge disorders [14], and doping [15,16]. Besides, the wave nature of phonon can also be utilized to modulate thermal properties of GNR, which can be realized in phononic metamaterials (PM) [3,17]. Based on local resonant hybridization, phonon transport is blocked, thus thermal conductivity can be dramatically reduced in PM, which is expected to remain excellent electric properties in the meanwhile. Previously, local resonant hybridization in silicon structures have been investigated in detail [3,17-19]. By introducing pillars on a plate, numerous local resonances take place. flat branches appear in the phonon dispersion, The local resonant hybridization significantly reduces the group velocities and consequently, the thermal conductivity [19].

However, the research on thermal transport properties of GNR-based nanostructures based on local resonant hybridization is relatively less. And, the development of GNR-based nanostructures with optimal thermal properties requires massive time and effort due to their huge parameter space. Thus, it is needed to obtain knowledge of optimal structures to guide the research by using an accelerated development process [4].

Recently, by combining material science and data science, machine learning algorithms have been widely applied in designing nanostructures and optimizing the target property [20,21]. In the field of heat conduction, the approach has been extended to the design of structures in atomic scale with optimized lattice thermal conductance or thermal conductivity [21-23]. These works have shown that such a method can considerably bridge the gap between structures and properties. Thus, the complex correlations among GNR-based nanostructures and their thermal properties based on phonon local resonant can be recognized utilizing machine learning algorithms, realizing structural optimization towards thermal transport properties of GNR within less property calculations.

In this work, thermal transport properties of GNRs are modulated by designing the pillars' arrangement. The optimization of thermal transport properties of GNRs is carried out based on the combination of Bayesian optimization (BO) and atomic Green's function (AGF), because ergodic evaluation for thermal properties of all the candidates is expensive [5,10]. Firstly, the effectiveness of the optimization method is demonstrated. Then, the optimization on GNRs with different length is performed. Lastly, the underlying physical mechanism is analyzed by the mode-analysis utilizing lattice dynamics. the effect of phonon scattering on the results is taken into account by performing nonequilibrium molecular dynamics (NEMD) simulations.

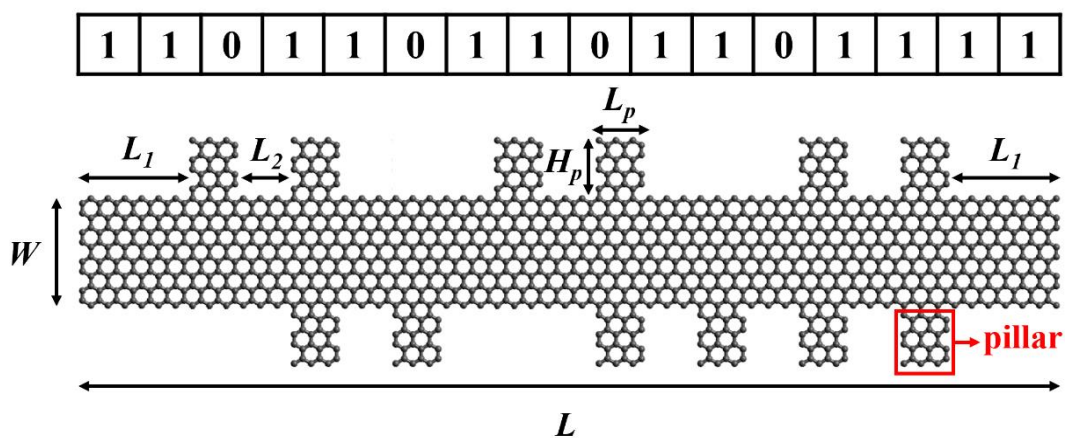

Fig. 1. Schematic picture of pillared GNR and corresponding binary flag.

## 2. Methods

Due to phonon resonance hybridization, the introduction of pillar, which is served as resonator, can have high impact on thermal transport. How to arrange the pillars to obtain the largest/ smallest thermal conductance? In this work, the arrangement of pillars on the GNR is designed to tune heat conduction in GNR based on the combination of AGF and BO. As shown in **Fig. 1**, the system is comprised of graphene nanoribbon and pillars on two sides, where pillars serve as resonators and the values of *W*, $L_1$, $L_2$, $L_P$, and $H_P$ are fixed to 1.58, 1.62, 0.87, 0.62 nm and 0.86 nm, respectively. In previous research, it has been observed that there are many flat bands in the phonon dispersion of GNRs with pillars in this size [16], which means that phonon local resonance takes place [3,19]. The thickness (*d*) of pillared GNRs is 0.334 nm with lattice constant (*a*) 0.1438 nm.

In the optimization part for thermal transport properties of pillared GNR, four elements are of basic requirement: the descriptor, evaluator, optimization method, and calculator. In this study, a binary flag is set as descriptors to describe the state of each pillar: "1" and "0" represent there is an occupied pillar or there is no pillar on the corresponding site, respectively. The thermal conductance (TC) is chosen as the evaluator for the quantitative evaluation of the performance of each candidate. Open-source Bayesian optimization library COMBO is employed to perform the optimization process [24], which has been successfully applied in recent work [4,23]. The details of Bayesian optimization can be found in the Supplemental Material (SM).

The AGF method [25,26] and NEMD simulations were utilized to calculate thermal transport properties of GNRs. Based on the Landauer formula [27], the value of TC ($\sigma$) at temperature *T* (300 K in this work) can be obtained by

$$\sigma = \frac{\hbar^2}{2\pi k_B T^2 S} \int \omega^2 \Xi(\omega) \frac{e^{\hbar\omega/k_B T}}{(e^{\hbar\omega/k_B T} - 1)^2} d\omega \qquad (1)$$

where $k_B$ is the Boltzmann constant, $\omega$ is the phonon frequency, $\Xi(\omega)$ is the phonon transmission function calculated by AGF method and *S* is the cross-sectional area. The details of transmission function calculation can be found in the SM. In order to take high-order interaction into account, further NEMD simulations are conducted for validation. The Large-scale Atomic/Molecular Massively Parallel Simulation (LAMMPS) package is used in the simulations [28-32]. The interatomic interactions are described by the optimized Tersoff potential, which has successfully reproduced the thermal transport properties of graphene [18,33-35]. The detailed parameters and specific simulation process are shown in the SM. Time step is set as 0.5 fs. Two Langevin thermostats [36] with a temperature difference of 20 K (310 K and 290K) are used to establish a temperature gradient along the longitudinal direction. The thermal conductivities shown here are averaged over five independent simulations with different initial conditions.

## 3. Results and Discussions

The optimization is divided to two steps: the first step is to testify the effectiveness of the proposed method by using a short enough pillared GNR, in which all of the candidates can be calculated in a short time; and the second is to apply this method to the optimization of several longer systems. In the first step, the GNR have 10 sites for pillars corresponding to 9.8 nm in length. So, the number of all the candidates is $2^{10}$(1024). This number is small enough, so that ergodic computation can be implemented to confirm the optimization results and the efficacy of the method. **Fig. 2(a) and 2(b)** exhibit the obtained optimal structures for maximum and minimum TC for GNR, respectively.

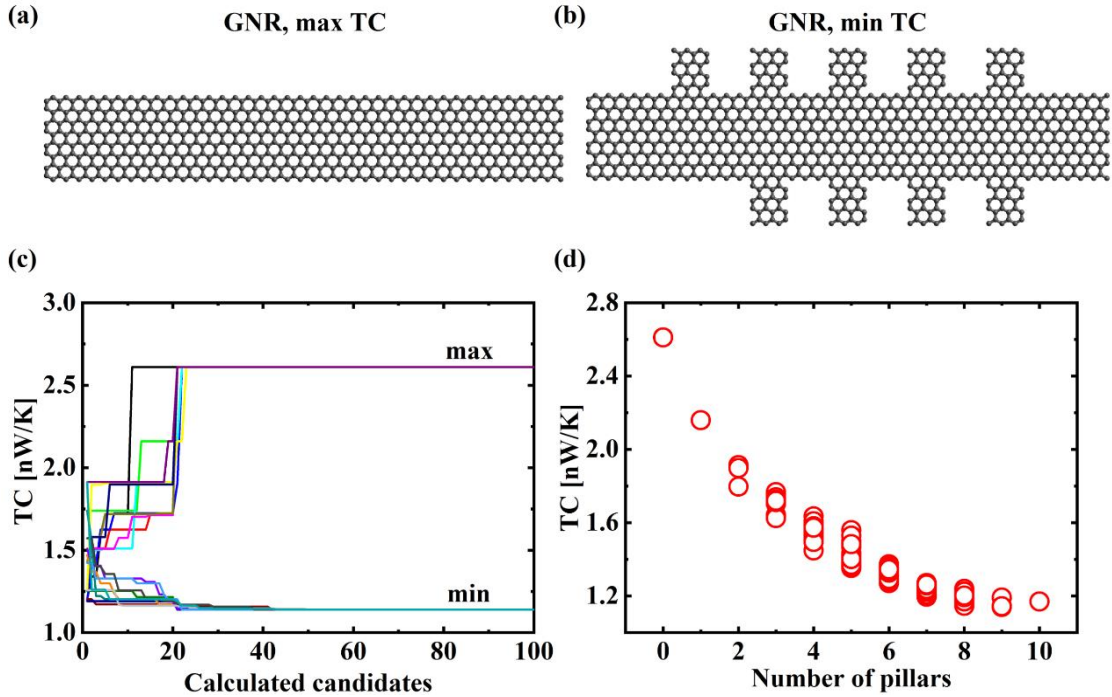

Fig. 2. (a)-(b) Optimal structures with the maximum and minimum thermal conductance for the system with the length of 9.8 nm. (c) The 10 optimization runs with different initial choices of candidates. (d) Thermal conductance versus number of pillars obtained from calculations of all the candidates.

The optimal structures exhibited here are the identical results from 10 independent optimization with different initial choices. The purpose of multi-optimization is to test the performance of Bayesian optimization algorithm. As shown in **Fig. 2(c)**, 10 optimizations converge within calculations of 43 candidates, which is only 4.2% of all candidates. As for the accuracy of the optimization, the TCs of all candidates are calculated to verify the optimal structures, and it is confirmed that the extremum of TC (maximum and minimum) and the corresponding structures are exactly consistent with those from BO.

The distributions of TCs of all candidates with different number of pillars is shown in **Fig. 2(d)**. The ratio of maximum to minimum TC is 2.3, which suggests that the TC is obviously dependent on the distribution of pillared nanostructure. Besides, for the structures with the same number of pillars, the TC values vary with the distribution of pillars, which indicates that phonon local resonant hybridization correlates with the site of resonators. Interestingly, the GNR with 10 sites fully occupied (1111111111) doesn't show the minimum thermal conductance, which is out of expectation. In a shorter system (with 8 sites), it is found that pillars enhance phonon local resonant hybridization so that the thermal conductance decreases, and when the 8 sites are fully occupied (11111111), the TC of GNR get minimum value (as shown in SM II). Since the pillars do not always make a negative effect on the thermal transport with the increase of length of GNR. It is moved on to second step and extend the optimization to several longer systems, where the number of sites are 12, 14, and 16, corresponding to 11.3, 12.8, and 14.3 nm in length, respectively. **Table 1** lists the range of TC values in various total length, all of the optimal structures, and the relative difference $\delta$ between the minimum TC value and the TC value of the GNR with fully occupied sites, which is defined as $\delta = (\sigma_{fo} - \sigma_{min})/\sigma_{min}$. As expected, the range of TCs is enlarged with the increase of the numbers of total candidates, where there is little much variation in the maximum TC, and a little reduction in the minimum TC. It can be seen that the structure with the maximum TC have no pillars on two sides in all cases, which is understandable because the pillar induces phonon local resonant, blocking heat conduction. As the length of GNR increases, the structure with the minimum TC always lacks one or two pillars, compared with the GNR with fully occupied sites, and exhibits the aperiodicity and asymmetry. Though the difference between these two structures is very small, there is always a nonnegligible gap of thermal transport properties among them, which also enlarge in the longer GNR without convergent tendency. To be emphasized, reasonably arranging the distribution of pillars on the GNR can further reduce the TC on the basis of phonon local resonant.

Table 1. Optimal thermal conductance and GNR-based nanostructures obtained through Bayesian optimization for different length.

| Length [nm] | TC [nW/K] | Optimal Structures [min/max] | $\delta$ |
| --- | --- | --- | --- |
| 9.8 | 1.14/2.61 | 1111101111/0000000000 | 3.0% |
| 11.3 | 1.07/2.62 | 111101101111/000000000000 | 6.5% |
| 12.8 | 1.03/2.63 | 11011111101111/00000000000000 | 9.7% |
| 14.3 | 0.99/2.63 | 0111111111110111/0000000000000000 | 12.1% |

Now that the structures with the minimum TC are identified, the mechanisms behind the smaller TC need to be investigated. Firstly, phonon dispersion relations of the GNR with fully occupied sites and the GNR with the minimum TC are compared. As a representative instance, the longest system (16 sites, 14.3 nm) is investigated in detail, where three structures are chosen for comparison: the GNR with the maximum TC (0000000000000000), the GNR with the minimum TC (0111111111110111), and the GNR with fully occupied sites (1111111111111111).

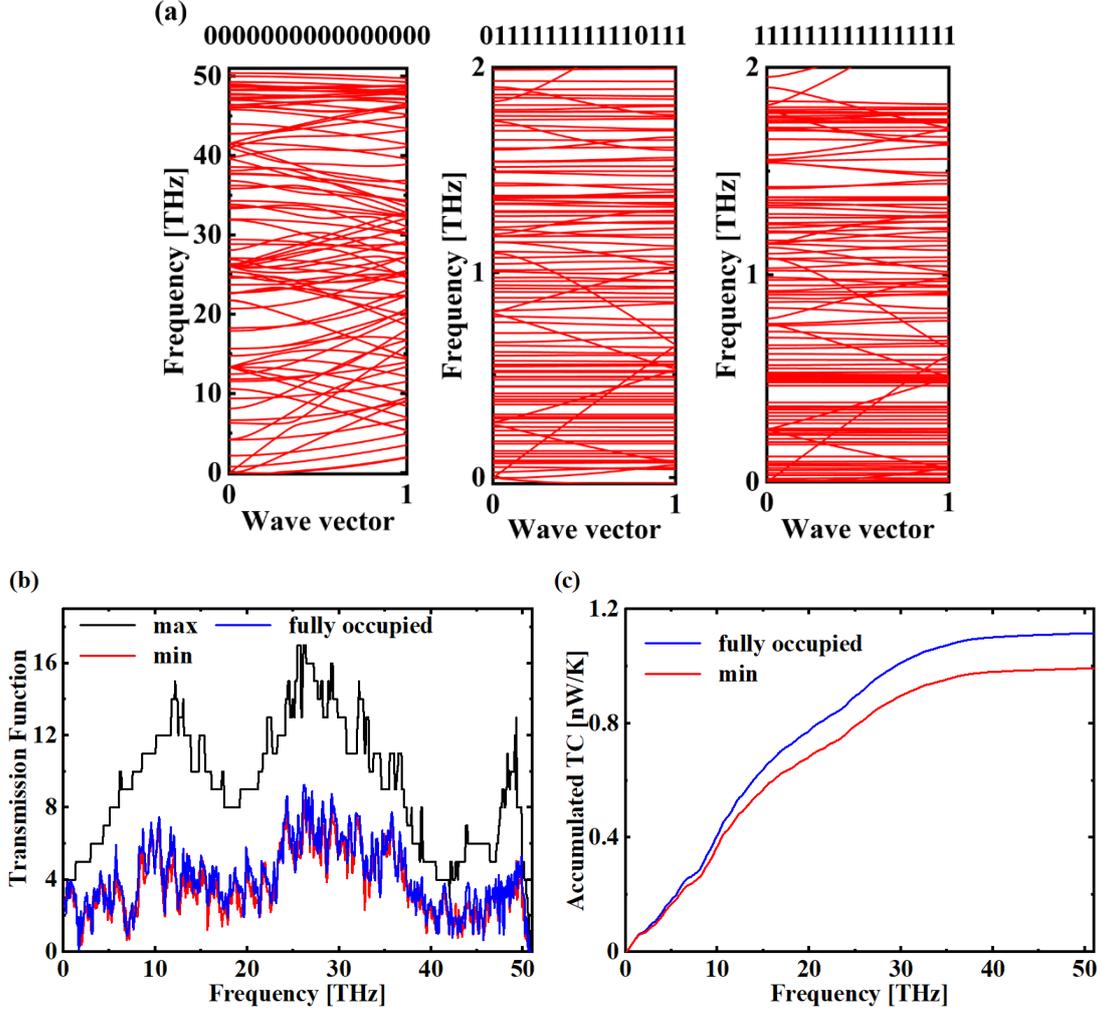

Fig. 3. The total length of the system in this case is fixed as 14.3 nm. (a) The phonon dispersion relationship of the GNRs with the maximum TC, the minimum TC, and fully occupied sites, respectively. (b) The phonon transmission function of these three structures. (c) The accumulated thermal conductance of the GNRs with the minimum TC and fully occupied sites.

The phonon dispersion relationship and phonon transmission function are shown in **Fig. 3(a)** and **3(b)**, respectively. Compared to the GNR (0000000000000000), the other two GNRs (0111111111110111 and 1111111111111111) exhibit many additional flat bands in the phonon dispersion, which means phonon local resonance at corresponding frequencies [3]. More resonant modes will induce more hybridization and impede phonon transport, which leads to the decrease of group velocities [16]. For the GNRs (0111111111110111) and (1111111111111111), although their TC values differ by 12.1%, the difference in their phonon dispersion is very small, which is not obvious in **Fig. 3(a)**. To better show the difference, the transmission function and accumulated thermal conductance of these three structures are calculated and shown in **Fig. 3(b)**. The transmission function of pristine GNR (0000000000000000) is much higher than those of latter two structures over a wide frequency range. More importantly, the GNR with the minimum TC exhibits the lower TC due to the lower transmission function in the frequency of 10-40 THz. This implies that it is by arranging the distribution of fixed-size pillars on the base-structure that phonon transport could be blocked in a wide frequency range, even though single pillar with fixed size can only hybridize some phonon modes at certain frequency.

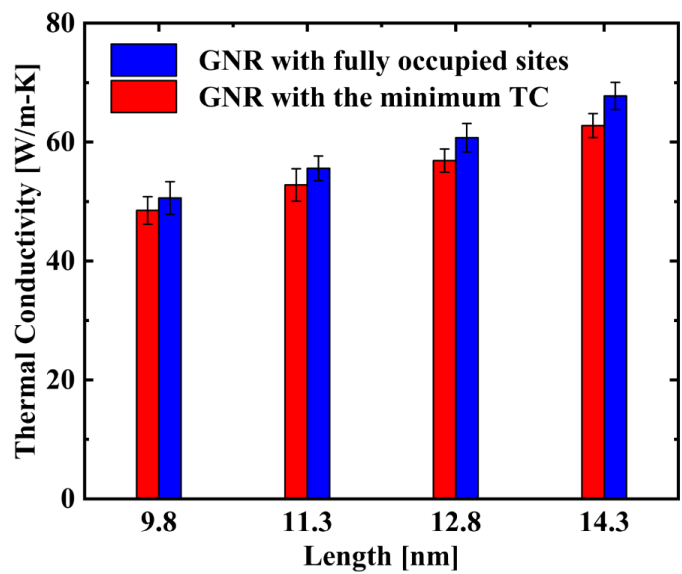

Fig. 4. Thermal conductivities of the GNRs with fully occupied sites and the minimum TC among different total length $L$ obtained from NEMD.

To further expand the above-discussed conclusions to more realistic systems, NEMD is utilized in consideration of phonon-phonon scattering. The thermal conductivities of the GNRs with the minimum TC in different total length $L$ are calculated, and compared with the thermal conductivities of corresponding GNRs with fully occupied sites. Besides, the thermal conductivities of GNRs with no pillars that always exhibit the highest TC among different length (9.8, 11.3, 12.8, and 14.3 nm) are also studied to test size effect. The thermal conductivity values of these structures are 228.13±11.88, 264.00±14.63, 314.73±16.68, 344.41±10.83 W/m-K, respectively.

Obviously, thermal conductivity is a function of length, which is because the phonon transport is largely affected by the boundary scattering, when the size of structures is comparable with the phonon mean free path [10]. As shown in **Fig. 4**, the GNR with the minimum TC also exhibits lower thermal conductivity compared with the GNR with fully occupied sites in different systems. Furthermore, the difference of thermal conductivities among these two structures also increase with the length of whole system. It is indicated that the introduction of phonon scattering does not impact the inferences based on phonon local resonance above.

## 4. Conclusions

In summary, we optimize thermal transport properties of pillared GNR based on phonon local resonance, by identifying the structures with the minimum or maximum TC. It is interesting that although the pillar can introduce phonon local resonant hybridization wave effect to block thermal transport, the GNR with fully occupied sites does not exhibit the lowest TC. This phenomenon becomes more obvious as the length of whole system increases. According to AGF calculation and phonon mode analysis, it is concluded that arranging the distribution of resonators on the base-structure could block phonon transport in a wide frequency range, even though the size of them doesn't change, which means the resonant modes remain the same. In the end, we perform NEMD to confirm the results above in consideration of phonon scattering, which is consistent with AGF calculations. This work offers a new perspective for manipulating phonon transport in GNR-based structures.

## Acknowledgements

This work is sponsored by the National Key Research and Development Project of China No. 2018YFE0127800, Fundamental Research Funds for the Central Universities No. 2019kfyRCPY045 and Program for HUST Academic Frontier Youth Team. We are grateful to Shiqian Hu, Meng An, Xiaoxiang Yu, Wentao Feng and Han Meng for useful discussions. The authors thank the National Supercomputing Center in Tianjin (NSCC-TJ) and the China Scientific Computing Grid (ScGrid) for providing assistance in computations.

## Declaration of Competing Interest

There are no conflicts of interest to declare.